\documentclass[11pt,aps,showkeys,floatfix,prd,preprintnumbers,superscriptaddress,groupedaddress]{revtex4}

\usepackage{epsf}
\usepackage{epsfig}
\usepackage{graphics}
\usepackage{graphicx}
\usepackage{amssymb}
\usepackage{mathtools}
\usepackage{color}
\usepackage{hyperref}
\usepackage{multirow}
\usepackage{feynmp-auto}
\usepackage[normalem]{ulem}
\usepackage{relsize}

\usepackage{amsmath, latexsym, amssymb, hyperref, graphicx, slashed, color, multirow}
\usepackage[12hr]{datetime} 
\definecolor{nicered}{rgb}{0.7,0.3,.3}
\definecolor{nicegreen}{rgb}{.1,.5,.1}
\definecolor{darkblue}{rgb}{0,.1,.9}
\hypersetup{colorlinks, citecolor=darkblue ,linkcolor=blue, urlcolor=blue}

\usepackage{cancel}

\def\lapp{\mathrel{\rlap{\raise.5ex\hbox{$<$}}
                    {\lower.5ex\hbox{$\sim$}}}}
\def\gapp{\mathrel{\rlap{\raise.5ex\hbox{$>$}}
                    {\lower.5ex\hbox{$\sim$}}}}

\newcommand{\beq}{\begin {equation}}  
\newcommand{\eeq}{\end   {equation}} 
\newcommand{\bea}{\begin {eqnarray}} 
\newcommand{\eea}{\end   {eqnarray}}  
\newcommand{\baa}{\begin {array}   } 
\newcommand{\eaa}{\end   {array}   }     
\newcommand{\bit}{\begin {itemize} }
\newcommand{\eit}{\end   {itemize} }
\newcommand{\be }{\begin {equation}} 
\newcommand{\ee }{\end   {equation}}

\def\be{\begin{equation}}
\def\ee{\end{equation}}
\def\bea{\begin{eqnarray}}
\def\eea{\end{eqnarray}}

\def\beq{\begin{equation}}
\def\eeq{\end{equation}}
\newcommand{\beqa}{\begin{eqnarray}} 
\newcommand{\eeqa}{\end{eqnarray}}
\newcommand{\barr}{\begin{array}}
\newcommand{\earr}{\end{array}}

\def\gs{\mathrel{
   \rlap{\raise 0.511ex \hbox{$>$}}{\lower 0.511ex \hbox{$\sim$}}}}
\def\ls{\mathrel{
   \rlap{\raise 0.511ex \hbox{$<$}}{\lower 0.511ex \hbox{$\sim$}}}}

\newcommand{ \ETmiss}{\cancel{E}_T}

\begin{document}
\preprint{HRI-RECAPP-2026-04}

\title{\Large Distinguishing Higgs portal and neutralino dark matter via vector boson fusion}

\author{Amit Chakraborty}\email{amit.c@srmap.edu.in} 
\affiliation{ Department of Physics, School of Engineering and Sciences,
	SRM University AP, Amaravati, Mangalagiri 522240, India}	
\author{Tathagata Ghosh} \email{tathagataghosh@hri.res.in }
\affiliation{Harish-Chandra Research Institute,  Chhatnag Road, Jhunsi, Prayagraj 211019, India}
\affiliation{Homi Bhabha National Institute, Training School Complex, Anushakti Nagar, Mumbai 400 094, India}
\author{Rafiqul Rahaman}\email{rafiqul@if.usp.br} 
\affiliation{Instituto de Física, Universidade de S\~ao Paulo, S\~ao Paulo, SP - 05580-090, Brazil}


\begin{abstract}

\section*{Abstract}
Understanding the nature of dark matter (DM) is a fundamental challenge in particle physics. In this paper, we investigate the potential of vector boson fusion (VBF) processes at the Large Hadron Collider (LHC) to demonstrate, as a proof of principle, the feasibility of distinguishing between different dark matter scenarios, focusing on Higgs portal DM (HPDM) and neutralino DM in the $2j + \ETmiss$ final state and exploiting the distinctive kinematic features of the VBF jets and the missing transverse energy. Our study reveals that the polarization of weak bosons in VBF plays a crucial role in shaping the transverse momentum distributions of the tagged jets, with the jets being less energetic in the transverse direction for the Higgs portal scenario compared to the neutralino scenario. In addition, the kinematic variables $\Delta\eta$ and $\Delta\phi$ exhibit characteristic differences between the Higgs portal and neutralino DM signals, providing significant discriminating power between these scenarios. We further apply a Kolmogorov--Smirnov test using linear discriminant analysis to quantify the distinguishability of the signals and find that the Higgs portal signals can be differentiated from neutralino DM signals with a C.L. exceeding $5\sigma$, thereby establishing the viability of collider-based discrimination between dark matter models.
\end{abstract}


\maketitle

\section{Introduction}
\label{sec:Intro}

The Standard Model (SM) of particle physics has been remarkably successful in describing the known elementary particles and their interactions. Nevertheless, it does not provide a viable candidate for dark matter (DM), whose existence is firmly established by a wide range of astrophysical and cosmological observations~\cite{Rubin:1970zza,Zwicky:1937zza,Zwicky:1933gu,Hayashi:2006kw,Clowe:2006eq,Hu:2001bc,WMAP:2006bqn,WMAP:2012nax,Planck:2018vyg}. This shortcoming has motivated numerous extensions of the SM that predict new particles capable of accounting for the observed DM relic abundance.

In this work, our goal is to investigate, as a proof of principle, whether different dark matter scenarios can be distinguished using collider signatures at the LHC, focusing on Higgs portal DM (HPDM) and neutralino DM as representative examples.
In the Higgs portal framework, the DM particle -- either scalar or fermionic -- interacts with SM fields exclusively via the Higgs boson. In contrast, neutralino DM arises in supersymmetric extensions of the SM, where each SM particle is accompanied by a superpartner with opposite spin statistics. In many supersymmetric models, the lightest supersymmetric particle (LSP), typically the neutralino ($\chi^0$), is stable and provides a compelling spin-$1/2$ DM candidate.

Although both Higgs portal and neutralino scenarios can successfully account for DM, it is crucial to develop experimental strategies capable of distinguishing between them. Such discrimination is essential for validating specific theoretical frameworks and  guiding the development of more complete theories beyond the SM.

At hadron colliders, DM searches primarily rely on signatures involving large missing transverse energy, such as mono-jet plus missing energy and vector boson fusion (VBF) processes leading to di-jet plus missing energy final states. The VBF channel, characterized by two forward jets separated by a large rapidity gap and substantial missing transverse momentum, offers a powerful and complementary probe of DM production mechanisms owing to its significantly lower backgrounds.
In this article, we explore the VBF production mechanism as a powerful probe to differentiate between  Higgs portal and neutralino DM at the future High-Luminosity LHC (HL-LHC). We focus on the $2j+\ETmiss$ final state, where two forward jets accompany large missing transverse energy. In the VBF process, two electroweak gauge bosons ($W$ or $Z$) fuse to produce  a pair of DM particles. Importantly, the polarization structure of the exchanged vector bosons differs between the Higgs portal and neutralino scenarios, leading to distinct kinematic features of the forward jets, particularly in their transverse momenta and angular distributions.

In the Higgs portal scenario, the relevant collider process can be written as
\begin{equation}
pp \;\to\; jj\,\bigl(V^{*}V^{*} \to h\bigr) \;\to\; jj\,\chi\bar{\chi},
\qquad V = W, Z,
\end{equation}
where $\chi$ denotes a scalar or fermionic DM particle. In the fermionic case, an additional singlet scalar mediator is introduced to provide a renormalizable realization of the Higgs portal and to obtain the observed thermal relic abundance~\cite{Baek:2011aa}. 

For neutralino DM, the corresponding VBF-induced processes are
\begin{equation}
pp \rightarrow 
\chi^0_1 \chi^0_1 jj \; + \;
\chi^0_1 \chi^\pm_1 jj \; + \;
\chi^\pm_1 \chi^\pm_1 jj \; + \;
\chi^+_1 \chi^-_1 jj ,
\end{equation}
where $\chi_1^0$ and $\chi_1^\pm$ denote the lightest neutralino and chargino, respectively. The charginos subsequently decay into the neutralino LSP through off-shell $W$ bosons, producing soft visible particles. In scenarios with a compressed electroweakino spectrum, these decay products typically fail reconstruction, such that all channels effectively lead to the same $2j+\ETmiss$ final state.

In this paper, we present a detailed collider analysis of VBF signatures for both Higgs portal and neutralino DM scenarios, following the invisible Higgs decay search strategy at the HL-LHC~\cite{CMS:2018tip,Cepeda:2019klc}. We exploit the kinematic properties of the VBF jets and missing transverse energy to discriminate between the two DM hypotheses using the Kolmogorov--Smirnov (KS) test~\cite{Knuth:10.5555/270146,pratt2012concepts} and Linear Discriminant Analysis (LDA)~\cite{tharwat:2017,ghojogh2019linear}.
We choose a set of representative benchmark points, discussed later, for the DM mass and associated model parameters such that a statistical significance of approximately $5\sigma$ can be achieved in distinguishing between the different DM scenarios. While these benchmark points are not required to reproduce the observed relic density, they are consistent with existing experimental constraints and serve as a proof of concept to demonstrate the discriminating power of VBF observables at the HL-LHC.

The remainder of this paper is organized as follows. In Section~\ref{sec:HPDM}, we introduce the Higgs portal DM framework, followed by a brief description of neutralino DM in Section~\ref{sec:mssm}. The kinematic features of the VBF jets are discussed in Section~\ref{sec:kin_feat_jet}. Our results are presented in Section~\ref{sec:results}, and we conclude with a summary and outlook in Section~\ref{sec:conclusion}.

\section{Simplified Model for Higgs portal DM}\label{sec:HPDM}
The Higgs portal DM model extends the SM by introducing a minimal set of additional particles and interactions. In this framework, DM interacts with the particles of the SM primarily through the Higgs boson. In the case of fermionic DM, an extra scalar particle is also included in the particle spectrum to facilitate the thermalization of the DM~\cite{Baek:2011aa}. Below, we provide a concise description of fermionic DM, followed by the scalar DM scenario.

\subsection{Higgs portal fermionic DM}\label{sec:HPFDM}
In the Higgs portal fermionic DM (HPF-DM) case, the SM is extended by adding a Dirac-like singlet fermionic DM candidate $\chi$, along with a singlet scalar $S$ that mediates the interaction with the SM. The fermion carries a non-trivial dark charge in order to distinguish it from a right-handed heavy neutrino appearing in other BSM models. The Lagrangian of the model can be written as~\cite{Baek:2011aa},
\begin{equation}
{\cal L} = {\cal L}_{\text{SM}} + {\cal L}_{\text{SFDM}},
\end{equation}
where
\begin{eqnarray}\label{eq:sfdm-lag}
{\cal L}_{\text{SFDM}} &=&  \overline{\chi} ( i \slashed{\partial} - m_\chi ) \chi + \frac{1}{2} \partial_\mu S \partial^\mu S - \frac{1}{2} m_0^2 S^2 \nonumber\\
&-& \lambda \, \overline{\chi} \chi S - \lambda_{HS} H^\dagger H S^2 - \mu_{HS} S H^\dagger H - \mu_0^3 S - \frac{\mu_S^\prime}{3} S^3 - \frac{\lambda_S}{4} S^4,
\end{eqnarray}
along with the SM Higgs potential
\begin{equation}\label{eq:VSM}
 {\cal L}_{\text{SM}} \supset V_H = -\frac{\mu^2}{2} H^\dagger H + \frac{\lambda_0}{4} (H^\dagger H)^2.
\end{equation}

After the scalar fields acquire non-zero vacuum expectation values (vevs), the fields can be expanded as
\begin{equation}\label{eq:vevs}
	H =
	\begin{pmatrix}
		G^+\\ (v_H + h + i G^0)/\sqrt{2}
	\end{pmatrix}, \quad
	S = v_S + s,
\end{equation}
where $v_H$ and $v_S$ are the vevs of $H$ and $S$, respectively, with $h$ and $s$ denoting the physical scalar fields. Here, $G^+$ and $G^0$ are the Goldstone bosons that are eaten by the $W^+$ and $Z$ gauge bosons, respectively. The two scalar fields mix to form the mass eigenstates, 
\begin{eqnarray}\label{eq:higgs-mixing}
h_1 &=& h \cos\alpha - s \sin\alpha, \nonumber\\
h_2 &=& h \sin\alpha + s \cos\alpha,
\end{eqnarray}
where $\alpha$ is the mixing angle that diagonalizes the Higgs mass matrix. We identify $h_1$ as the observed $125$~GeV Higgs boson. The mixing between $h$ and $s$ leads to a universal suppression of the Higgs signal strengths at the LHC, independent of the production and decay channels~\cite{Baek:2011aa}.

The relevant interaction terms are given by
\begin{eqnarray}\label{eq:sfdm-Lint}
{\cal L}_{\rm int} & = &  - ( h_1 \cos\alpha + h_2 \sin\alpha ) \left[ \sum_f \frac{m_f}{v_H} 
 \overline{f} f  - \frac{2 m_W^2}{v_H} W_\mu^+ W^{-\mu} - 
 \frac{m_Z^2}{v_H} Z_\mu Z^\mu   \right]   \nonumber \\
 & + &  \lambda ( h_1 \sin\alpha - h_2 \cos\alpha ) \overline{\chi} \chi .
\end{eqnarray}

The quartic couplings can be expressed in terms of the Higgs mass parameters and the mixing angle as
\begin{eqnarray}\label{eq:lam-vs-higgs-mass}
\lambda_{HS} &=& \dfrac{m_{hs}^2 - \mu_{HS} v_H}{v_H v_S}, \nonumber\\
\lambda_S &=& \dfrac{m_{ss}^2 + \mu_0^3/v_S - \mu_S^\prime v_S - \mu_{HS} v_H^2/(2v_S)}{2v_S^2},
\end{eqnarray}
where $m_{hs}$ and $m_{ss}$ appear in the Higgs mass matrix
\begin{equation}\label{eq:Higss-mass-matrix}
M_H^2 =
\begin{pmatrix}
    m_{hh}^2 & m_{hs}^2 \\
    m_{hs}^2 & m_{ss}^2
\end{pmatrix}.
\end{equation}
The components of the matrix are given by~\cite{Profumo:2007wc}
\begin{eqnarray}\label{eq:SFDM-massmatrix}
m_{hh}^2 &=& 2 \lambda_0 v_H^2, \nonumber\\
m_{ss}^2 &=& \mu_S v_S + 2 \lambda_S v_S^2 - \frac{\mu_{HS} v_H^2}{2 v_S}, \nonumber\\
m_{hs}^2 &=& 2 ( \mu_{HS} + 2 \lambda_{HS} v_S ) v_H.
\end{eqnarray}

The mixing angle $\alpha$ and the mass eigenvalues are given by
\begin{eqnarray}
\tan\alpha &=& \frac{y}{1 + \sqrt{1 + y^2}}, \nonumber\\
m_{h_1,h_2}^2 &=& \frac{m_{hh}^2 + m_{ss}^2}{2} \pm \frac{m_{hh}^2 - m_{ss}^2}{2} \sqrt{1 + y^2},
\end{eqnarray}
where
\begin{equation}
y = \frac{m_{hs}^2}{m_{hh}^2 - m_{ss}^2}.
\end{equation}

The free parameters of the model can now be chosen as
\begin{equation}\label{eq:SFDM-free-param}
m_{h_2},~\alpha,~v_S,~\mu_0,~\mu_S^\prime,~\mu_{HS},~m_{\chi},~\lambda.
\end{equation}
We fix $v_S$, $\mu_0$, $\mu_{HS}$, and $\mu_S^\prime$ to be equal to $m_{h_2}$ for completeness. These parameters have no impact on the collider phenomenology discussed in this work. Using the shift symmetry of the scalar potential, the linear term in $S$ in Eq.~(\ref{eq:sfdm-lag}) can be removed and $\mu_0$ absorbed into other parameters, with the remaining potential parameters modified accordingly.

For the collider phenomenology of interest, the three relevant BSM parameters are the heavy scalar mass $m_{h_2}$, the mixing angle $\cos\alpha$, and the DM--scalar coupling $\lambda$. We tune $\lambda$ such that $\Gamma_{h_2}/m_{h_2} < 0.2$, while satisfying the perturbative  constraint $\lambda \leq \sqrt{4\pi}$. Constraints on the mixing angle arise from Higgs signal strength measurements at the LHC. For instance, ATLAS~\cite{ATLAS:2015ciy,ATLAS:2016neq} constrains the mixing angle to $\sin\alpha < 0.33$ ($\cos\alpha > 0.94$). Stronger constraints on $\sin\alpha$ in the range $\sim[0.3,0.2]$ originate from next-to-leading order corrections to the $W$-boson mass~\cite{Robens:2015gla} for $250~\text{GeV} \lesssim m_{h_2} \lesssim 800~\text{GeV}$ (see Fig.~3 of Ref.~\cite{Robens:2015gla}). We therefore choose a conservative benchmark point with $m_{\chi} = 130$~GeV, $m_{h_2} = 275$~GeV, $\cos\alpha = 0.95$, and $\lambda = 3$. In this setup, we exploit resonant $h_2$ production via VBF to enhance the $pp \rightarrow \chi \chi j j$ production cross section.

\subsection{Higgs portal scalar DM}
A simplified model for Higgs portal scalar DM (HPS-DM),  can be obtained by removing the fermion $\chi$ and imposing a $Z_2$ symmetry on $S$ to ensure its stability in the Lagrangian of Eq.~(\ref{eq:sfdm-lag})~\cite{Kanemura:2010sh}. In this case, only $\lambda_{HS}$ and $\lambda_S$ are non-zero, while all other BSM parameters vanish. The Lagrangian is given by
\begin{eqnarray}
{\cal L} = {\cal L}_{\rm SM} + \frac{1}{2} \partial_\mu S \partial^\mu S - \frac{1}{2} m_0^2 S^2 - \lambda_{HS} H^\dagger H S^2 - \frac{\lambda_S}{4} S^4.
\end{eqnarray}
After electroweak symmetry breaking, the scalar DM mass is given by
\begin{equation}
 m_S^2 = m_0^2 + \lambda_{HS} v_H^2.
\end{equation}
The interaction of scalar DM with the SM Higgs is
\begin{equation}
{\cal L}_{\rm int} = 2 \lambda_{HS} v_H h S^2.
\end{equation}

For this scenario, we choose a benchmark similar to the HPF-DM case, namely a DM mass $m_S = 130$~GeV and $\lambda_{HS} = 3$.

\section{Minimal supersymmetric SM}
\label{sec:mssm}

Supersymmetry (SUSY) extends the SM by associating a superpartner with opposite spin statistics to each SM particle. The Minimal Supersymmetric SM (MSSM) represents the simplest phenomenologically viable realization, requiring two Higgs doublets with opposite hypercharge to ensure anomaly cancellation. Supersymmetry stabilizes the electroweak scale and is assumed to be softly broken, leading to superpartner masses above current experimental limits.

A discrete symmetry known as $R$-parity, defined as $R=(-1)^{3B-L+2s}$, where $B$ is the baryon number, $L$ is the lepton number, and $s$ is the  spin, is imposed to prevent rapid proton decay. Under $R$-parity, SM particles are even while superpartners are odd, rendering the LSP stable. In the MSSM, a neutral LSP with electroweak interactions, namely the lightest neutralino, provides a well-motivated DM candidate.

The Higgs sector consists of two scalar doublets,
\begin{equation}
H_1=\begin{pmatrix} H_1^0 \\ H_1^- \end{pmatrix}, \qquad
H_2=\begin{pmatrix} H_2^+ \\ H_2^0 \end{pmatrix},
\end{equation}
with vacuum expectation values $v_1$ and $v_2$, satisfying $v=\sqrt{v_1^2+v_2^2}=246~\mathrm{GeV}$ and $\tan\beta=v_2/v_1$. The corresponding fermionic superpartners are the higgsinos $\tilde{H}_{1,2}$, while the electroweak gaugino sector contains the bino $\tilde{B}$ and the winos $\lambda^{\pm},\lambda^3$.

After electroweak symmetry breaking, the charged winos and higgsinos mix to form two charginos $\chi_{1,2}^\pm$, while the neutral bino, neutral wino, and neutral higgsinos mix to form four Majorana neutralinos $\chi_i^0$. Their masses and compositions are governed by the soft gaugino masses $M_1$ and $M_2$, the higgsino mass parameter $\mu$, and $\tan\beta$. The lightest neutralino $\chi_1^0$ is identified as the LSP and constitutes the DM candidate in this work. Since our analysis focuses on a compressed electroweakino spectrum, production channels involving charginos also contribute significantly to the signal. The charginos decay promptly into the neutralino LSP via off-shell $W$ bosons, yielding only soft visible decay products. Owing to the small mass splittings, these particles typically fall below reconstruction thresholds, such that all electroweakino production channels effectively manifest as the same 
$2j+\ETmiss$ final state at the detector level.

In this work, we focus on scenarios with a compressed electroweakino spectrum, where all squarks, sleptons, and gluinos are decoupled with masses above $5~\mathrm{TeV}$, and only the lightest electroweakinos remain near the electroweak scale. We consider two representative benchmark scenarios for the neutralino LSP DM, corresponding to different dominant compositions: a bino–wino mixture and a higgsino-like state. In both cases, the lightest neutralinos are kept light while the heavier electroweakinos are decoupled. These benchmark choices are consistent with current compressed SUSY searches at the LHC~\cite{ATLAS:2019lng,ATLAS:2021moa,ATLAS:2024umc}, and are described below.

\paragraph{Wino/Bino DM : }\label{sec:wino-dm}
In this case, we consider a lightest neutralino ($\chi_1^0$) mass of $130$ GeV having $98\%$ bino LSP and $98\%$ wino next to LSP (NLSP) component having $\Delta M = m_{\chi_1^\pm}-m_{\chi_1^0}=2$ GeV and $m_{\chi_2^0}\equiv m_{\chi_1^+}$.

\paragraph{Higgsino DM :  }\label{sec:higgsino-dm}
For Higgsino-like DM, we consider a $100\%$ Higgsino LSP with the mass of $\chi_1^0$ to be $130$ GeV with $\Delta M =2$ GeV.
This higgsino benchmark is not phenomenologically relevant in this context, as its significance in our analysis is rather low, described later, and it is already marginally excluded by Refs.~\cite{ATLAS:2024umc,ATLAS:2025lhc}. Nevertheless, we retain this benchmark for illustrative purposes, in order to highlight the differences among various types of DM candidates.

\section{Kinematic feature of VBF jets}\label{sec:kin_feat_jet}
\begin{figure}[!htp]
\includegraphics[width=0.49\textwidth]{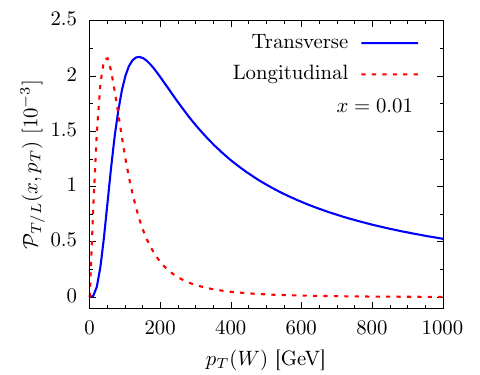}
\includegraphics[width=0.49\textwidth]{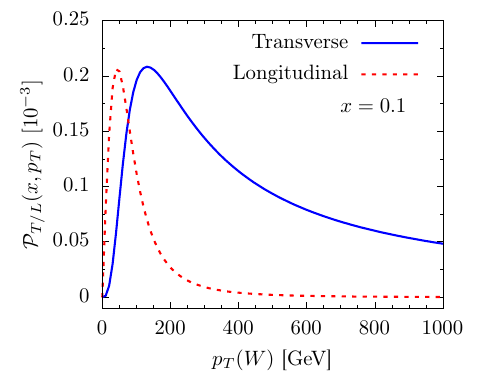}
\caption{Structure functions  for longitudinally ($W_L$) and transversely ($W_T$) polarized $W$ bosons in the EWA, shown as functions of $p_T$ for $x=0.01$ (left panel) and $x=0.1$ (right panel).}
\label{fig:W-pdf_EWA}
\end{figure}
The polarization of weak gauge bosons plays a crucial role in determining the kinematic properties of the recoiling jets in a vector boson fusion (VBF) process, leading to distinct features for the Higgs portal and wino-like DM scenarios. Although direct $W$-boson beams are not available, at sufficiently high energies the radiated weak bosons can be treated as partons emitted from the incoming quarks. This framework is known as the Effective $W$ Approximation (EWA)~\cite{Dawson:1984gx,Kane:1984bb,Borel:2012by}.  

Within the EWA and in the high-energy limit $\sqrt{\hat{s}} \gg M_W$, the probability of finding a $W$ boson radiated from an incoming quark with longitudinal momentum fraction $x$  and transverse momentum $p_T$ is given by~\cite{Borel:2012by}
\begin{eqnarray}
\mathcal{P}_T(x,p_T) &=& \dfrac{g^2}{16\pi^2}\,\dfrac{1+(1-x)^2}{x}\,
\dfrac{p_T^3}{\big((1-x)M_W^2+p_T^2\big)^2}\,,\nonumber\\
\mathcal{P}_L(x,p_T) &=& \dfrac{g^2}{16\pi^2}\,\dfrac{(1-x)^2}{x}\,
\dfrac{2p_T M_W^2}{\big((1-x)M_W^2+p_T^2\big)^2}\,,
\end{eqnarray}
where the subscripts $L$ and $T$ denote longitudinal and transverse polarizations of the $W$ boson, respectively. The $x$- and $p_T$-dependent probabilities can be interpreted as structure functions of the $W$ boson inside the emitting jet, exhibiting distinct behaviors for longitudinal and transverse polarizations. 

From these expressions, it is evident that a VBF-tagged jet recoiling against a transversely polarized $W$ tends to be harder than one recoiling against a longitudinally polarized $W$. This behavior is illustrated in Fig.~\ref{fig:W-pdf_EWA}, which shows the $p_T$ distributions of the $W$ boson for representative values $x=0.01$ (left panel) and $x=0.1$ (right panel). Consequently, jets produced in $W_LW_L$ scattering are expected to be more forward than those arising from $W_TW_T$ scattering. Although the EWA is not quantitatively reliable over the entire $\sqrt{\hat{s}}$ range relevant for our analysis, it nevertheless captures the essential qualitative features of the underlying dynamics.

In the Higgs portal scenario, the VBF jets are therefore expected to be relatively softer, as the Higgs boson couples predominantly to the longitudinal components of the weak gauge bosons in accordance with the Goldstone boson equivalence theorem~\cite{Cornwall:1974km}. In contrast, for a pure wino-like LSP, DM pair production proceeds dominantly through $V_TV_T$ scattering, with only a small contribution from $V_LV_L$ modes due to the suppressed couplings involving the Goldstone bosons $G^\pm$ and the electroweakinos $\chi^\pm$ and $\chi^0$~\cite{Kuroda:1999ks}. As a result, the associated VBF jets in the wino-like scenario are typically harder and less forward compared to those in the Higgs portal case.

\section{Results}
\label{sec:results}
We estimate the production cross sections of the processes discussed in different DM models with the chosen parameter values at leading order (LO) in QCD, and compute the expected number of events for the four signal scenarios, namely HPF-DM, HPS-DM, Wino-DM, and Higgsino-DM, using the cut strategy for the Higgs  to invisible decay search at the High-Luminosity LHC (HL-LHC)~\cite{CMS:2016dhk}. The analysis is performed at a center-of-mass energy of $\sqrt{s}=14$ TeV with an integrated luminosity of $3$ ab$^{-1}$.
 The kinematic cuts for the VBF $h\to$ invisible analysis are summarized in Table~\ref{tab:vbf_cuts}. We use \textsc{MadGraph5\_aMC@NLO}~\cite{Alwall:2014hca} for event generation, and \textsc{Pythia8}~\cite{Sjostrand:2014zea} for parton showering and hadronization, followed by fast detector simulation in \textsc{Delphes}~\cite{deFavereau:2013fsa} using the default parameter settings.
\begin{table}[b!]
    \caption{VBF and other kinematic cuts for  the Higgs invisible decay analysis~\cite{CMS:2018tip,Cepeda:2019klc}.}
	\renewcommand{\arraystretch}{1.50}
	\centering	
    \begin{tabular}{|c|l|}\hline
cut\_1 & Basic cuts: $p_{T_{j_{1,2}}} > 20$ GeV, $|\eta_{j_{1,2}}|<5$ \\ \hline
cut\_2 & $\ell$-veto, $\tau_h$-veto, $b$-veto \\ \hline
cut\_3 & VBF cuts: $p_{T_{j_{1 (2)}}} > 80 \, (40)$ GeV, $M_{j_1 j_2} > 2500$ GeV, $|\Delta \eta_{j_1 j_2}| > 4.0$, $\eta_{j_1} \eta_{j_2} < 0$, $\Delta \phi_{j_1 j_2} > 1.8$\\ \hline
cut\_4 & $\cancel{E}_T > 190$ GeV \\ \hline
cut\_5 & $\Delta \phi_{j \cancel{E}_T} > 0.5$  \\ \hline
    \end{tabular}
    \label{tab:vbf_cuts}
\end{table}

\begin{table}
	\caption{\label{tab:cutflow-table} Production cross-sections, number of events, and signal significances after all cuts (Table~\ref{tab:vbf_cuts}) for the four types of DM signals at $\sqrt{s}=14$ TeV and an integrated luminosity of ${\cal L}=3$ ab$^{-1}$.}
	\renewcommand{\arraystretch}{1.50}
	\centering	
 \begin{scriptsize}
	\begin{tabular}{|c|c|c|c|c|}\hline
		& HPF-DM & HPS-DM  & Wino-DM & Higgsino-DM \\ \hline
  Benchmark & \begin{tabular}{c}
 $m_{\chi} = 130$ GeV,\\ $m_{h_2} = 275$ GeV,\\ $\cos \alpha = 0.95$, \\ $\lambda = 3$
  \end{tabular} & 
  \begin{tabular}{c}
  $m_S=130$ GeV,\\ $\lambda_{HS}=3$
   \end{tabular}
&
 \begin{tabular}{c}
 $98\%$ bino LSP,\\ $98\%$ wino NLSP,\\
  $m_{\chi_1^0}=130$ GeV,\\ $\Delta M =2$ GeV
 \end{tabular} & 
 \begin{tabular}{c}
 $100\%$ Higgsino,\\ $m_{\chi_1^0}=130$ GeV,\\ $\Delta M =2$ GeV
  \end{tabular} \\ \hline
 Production process & \multicolumn{2}{c|}{$\chi\chi jj$}   &\multicolumn{2}{|c|}{$\chi^0_1 \chi^0_1 jj  + \chi^0_1 \chi^\pm_1 jj  + 
\chi^\pm_1 \chi^\pm_1 jj  + \chi^+_1 \chi^-_1 jj$} \\ \hline
Generation-level cross-section (fb) & $96.26$& $68.58$ &$108.48$ & $37.52$ \\ \hline
Cross-section after all cuts (fb)& $0.43$  &$0.49$ &$0.47$ & $0.09$ \\ \hline
Number of events after all cuts & $1300$  &$1466$ &$1420$ & $280$ \\ \hline
Total background events after all cuts~\cite{CMS:2018tip,Cepeda:2019klc}& \multicolumn{4}{c|}{64419} \\ \hline
Significance ($S/\sqrt{B}$) & $5.07$ & $5.71$ & $5.54$ & $1.10$\\ \hline
	\end{tabular}
  \end{scriptsize}
\end{table}

The generation-level production cross-sections, cross-sections after VBF and other kinematic cuts (see Table~\ref{tab:vbf_cuts}), and the corresponding number of events for an integrated luminosity of $3$ ab$^{-1}$ at the $14$ TeV LHC are presented in Table~\ref{tab:cutflow-table}. The estimated signal significances ($S/\sqrt{B}$) are shown in the last row of Table~\ref{tab:cutflow-table}, using the expected total background yields at $3$ ab$^{-1}$ taken from Refs.~\cite{CMS:2018tip,Cepeda:2019klc}.

In the HPDM case, the final state is $\chi \bar{\chi} j j$. 
For the HPF-DM scenario, and for the chosen benchmark point listed in Table~\ref{tab:cutflow-table}, the production cross-section for $pp \rightarrow \chi \chi j j$ is $96.26$ fb at $\sqrt{s} = 14$ TeV. Using the CMS VBF cuts from the $h \rightarrow \text{invisible}$ analysis~\cite{CMS:2016dhk}, we estimate a signal significance of approximately $5\sigma$ for an integrated luminosity of $3$ ab$^{-1}$. 
The significance can be further improved by lowering the values of $m_{h_2}$ and $\cos\alpha$, which would enhance the production cross-section. 
In the case of the HPS-DM scenario, a production cross-section of $68.58$ fb is obtained for the chosen benchmark point, leading to a significance of $5.71\sigma$ after applying the VBF cuts.

For neutralino DM in the MSSM, the relevant final states are $\chi^\pm\chi^\pm jj+\chi^+\chi^-jj+\chi^\pm\chi^0jj+\chi^0\chi^0jj$. 
In the Wino-DM case, the total production cross-section is $108.48$ fb, resulting in a signal significance exceeding $5\sigma$ after VBF cuts for an integrated luminosity of $3$ ab$^{-1}$. 
In contrast, for the Higgsino-DM scenario, the production cross-section is comparatively smaller ($37.52$ fb), leading to a reduced signal significance of approximately $1\sigma$. 
We note that systematic uncertainties have not been included in this analysis, and the significances quoted in Table~\ref{tab:cutflow-table} will be diluted once systematic effects are taken into account.

\begin{figure}
	\centering
	\includegraphics[width=0.45\textwidth]{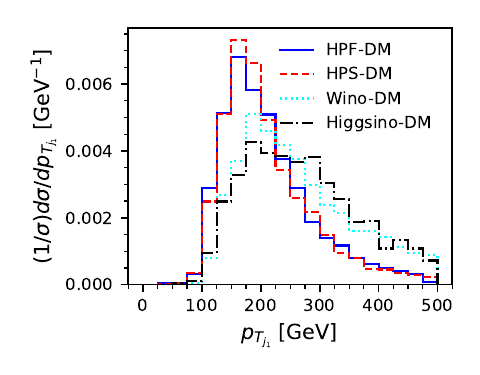}
	\includegraphics[width=0.45\textwidth]{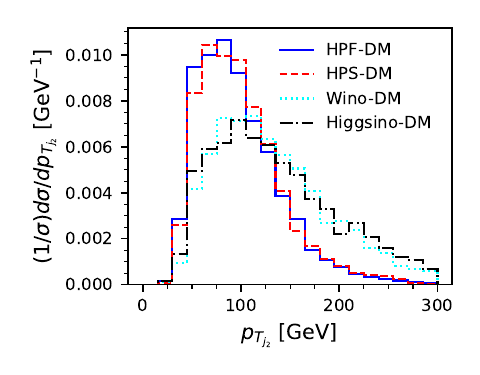}
	\includegraphics[width=0.45\textwidth]{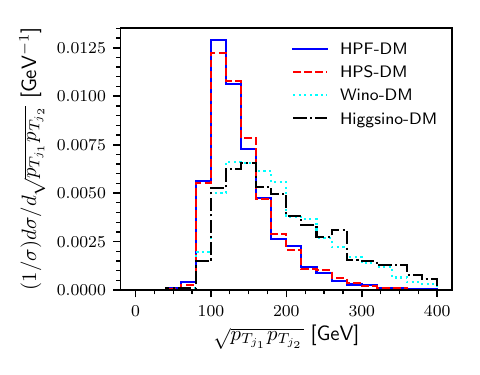}
	\includegraphics[width=0.45\textwidth]{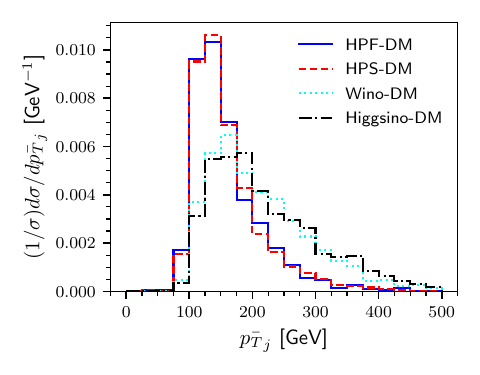}
	\includegraphics[width=0.45\textwidth]{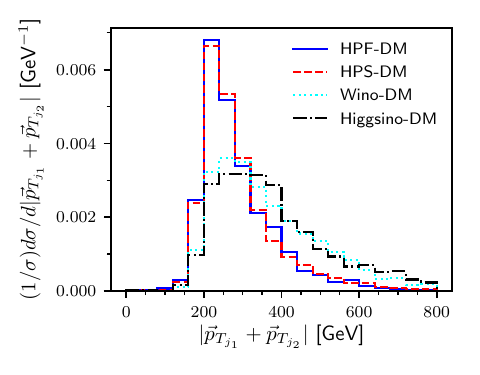}
	\includegraphics[width=0.45\textwidth]{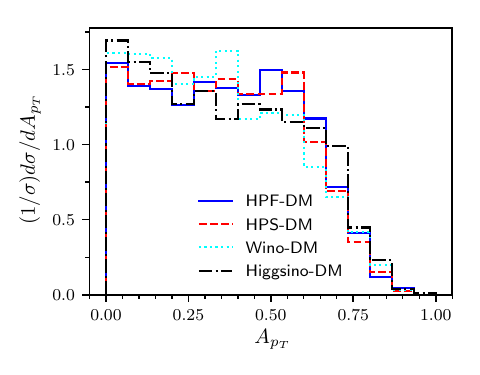}
	\includegraphics[width=0.45\textwidth]{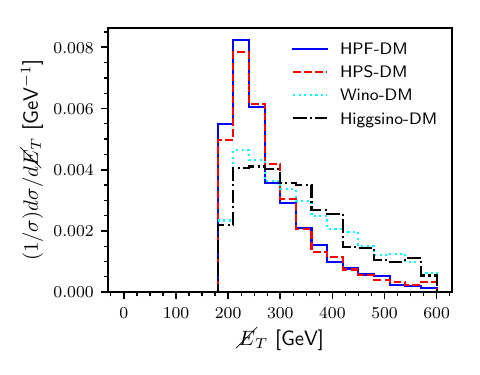}
	\includegraphics[width=0.45\textwidth]{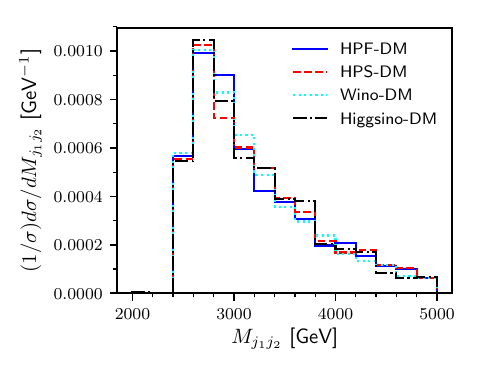}
	\caption{\label{fig:Dist-normal-1} Normalized distributions of the transverse momentum ($p_T$) of the VBF jets and their combinations, missing transverse energy ($\cancel{E}_T$), and dijet invariant mass ($M_{j_1j_2}$) for the four signal scenarios.
}
\end{figure}
\begin{figure}
	\centering
	\includegraphics[width=0.45\textwidth]{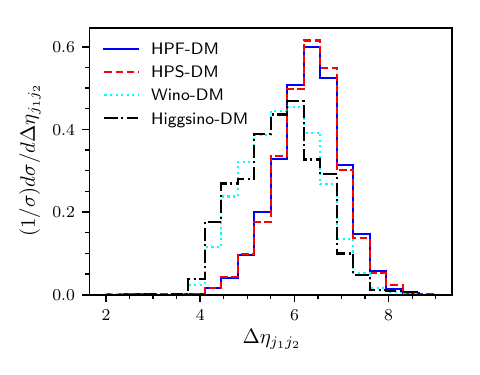}
	\includegraphics[width=0.45\textwidth]{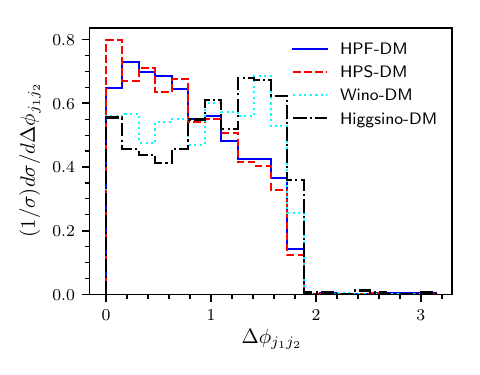}
	\includegraphics[width=0.45\textwidth]{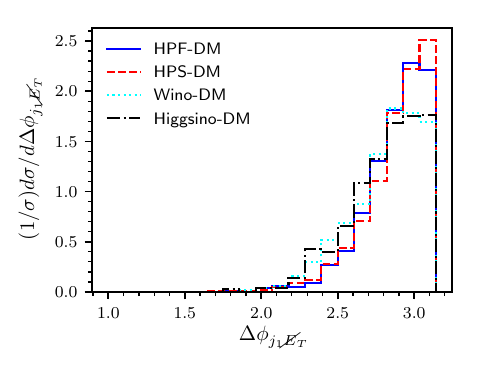}
	\includegraphics[width=0.45\textwidth]{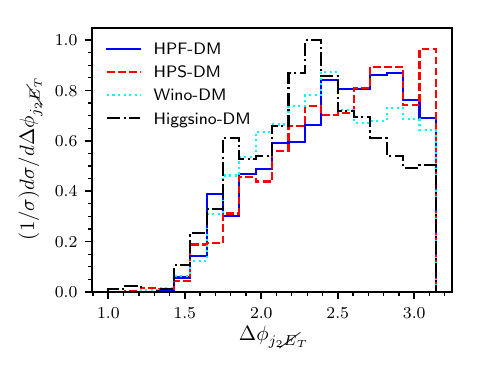}
	\caption{\label{fig:Dist-normal-2} Normalized distributions for $\Delta\phi$, $\Delta\eta$, and angular correlations among the VBF jets and $\cancel{E}_T$ for the four signal scenarios.}
\end{figure}

We now proceed to distinguish between the four DM signal scenarios using different kinematic variables, guided by the differences in VBF jet kinematics discussed in Sec.~\ref{sec:kin_feat_jet}. 
For a final state with two jets and missing transverse energy, the natural kinematic variables are:
\begin{itemize}
\item $p_{T_{j_{1,2}}}$: transverse momenta of the two jets,
\item $M_{j_1j_2}$: invariant mass of the two jets,
\item $\eta_{j_{1,2}}$: pseudorapidity of the two jets,
\item $|\Delta \eta_{j_1j_2}|$: absolute pseudorapidity separation of the two jets,
\item $\Delta \phi_{j_1j_2}$: azimuthal angular separation of the two jets,
\item $\cancel{E}_T$: missing transverse energy, and 
\item $\Delta \phi_{j_{1,2}\cancel{E}_T}$: azimuthal separation between the jets and the missing transverse momentum.
\end{itemize}

Additional variables can be constructed from the jet transverse momenta:
\begin{itemize}
\item $\bar{p}_T = \dfrac{p_{T_{j_{1}}}+p_{T_{j_{2}}}}{2}$: average transverse momentum,
\item $\sqrt{p_{T_{j_{1}}}  p_{T_{j_{2}}}}$: geometric mean of the jet transverse momenta,
\item $|\vec{p}_{T_{j_{1}}} + \vec{p}_{T_{j_{2}}}|$: magnitude of the vector sum of jet transverse momenta, and 
\item $A_{p_T} = \dfrac{p_{T_{j_{1}}} - p_{T_{j_{2}}}}{p_{T_{j_{1}}} + p_{T_{j_{2}}}}$: transverse momentum asymmetry between the two jets.
\end{itemize}

Normalized distributions of these variables, both independent and derived, are shown in Figs.~\ref{fig:Dist-normal-1} and~\ref{fig:Dist-normal-2} after applying the VBF cuts listed in Table~\ref{tab:vbf_cuts}. 
The distributions do not include the $\cancel{E}_T$ cut (cut\_4) and the $\Delta \phi_{j \cancel{E}_T}$ cut (cut\_5) due to limited Monte Carlo  statistics. 
As discussed in Sec.~\ref{sec:kin_feat_jet}, the VBF jets in the Higgs portal scenarios are softer compared to those in the Wino and Higgsino DM cases, as evident from Fig.~\ref{fig:Dist-normal-1} (top row). 
The $\cancel{E}_T$ distributions and other $p_T$-derived variables exhibit similar behavior. 
The invariant mass distributions show no significant differences among the signals.

Among the distributions shown in Fig.~\ref{fig:Dist-normal-2}, the $\Delta\eta$ and $\Delta\phi$ distributions of the two jets exhibit noticeable differences between the Higgs portal and Wino/Higgsino DM scenarios. 
In particular, the $\Delta\phi$ distribution peaks near zero for the Higgs portal case, while it peaks near $\pi/2$ for the SUSY scenarios. This motivates the construction of the following asymmetry:
\begin{equation}\label{eq:dphi-asym}
	A_{\Delta\phi} = \dfrac{\sigma(\Delta\phi_{j_1j_2}>\pi/3)-\sigma(\Delta\phi_{j_1j_2}<\pi/3)}{\sigma(\Delta\phi_{j_1j_2}>\pi/3)+\sigma(\Delta\phi_{j_1j_2}<\pi/3)}.
\end{equation}
The resulting asymmetry values are $-0.18$, $-0.20$, $+0.04$, and $+0.15$ for the HPF-DM, HPS-DM, Wino-DM, and Higgsino-DM scenarios, respectively. 
Thus, the asymmetry is negative for the Higgs portal cases and positive for the SUSY scenarios. 
The $\Delta\phi$ distributions between the jets and the missing transverse momentum do not show significant differences among the signals. There is essentially no distinction between the two Higgs portal scenarios, while a small difference is observed between the Wino and Higgsino cases.

With these qualitative features of the kinematic variables in hand, we proceed in the next subsection to quantify the differences among the signal scenarios using the Kolmogorov--Smirnov test.

\subsection{Distinguishing the DM signals using the Kolmogorov--Smirnov test}
Here, we estimate quantitative differences among the four signal scenarios using  LDA~\cite{tharwat:2017,ghojogh2019linear}, followed by the Kolmogorov--Smirnov (KS) test~\cite{Knuth:10.5555/270146,pratt2012concepts}. 

In the KS test, two hypotheses $H_0$ and $H_1$ are said to be distinguishable at a confidence level (C.L.) of $100\times(1-\alpha)\%$ if
\begin{equation}
D_{\rm max} = \left| CDF_N(H_0) - CDF_N(H_1) \right|_{\rm max} > D_\alpha ,
\end{equation}
where $D_\alpha = c_\alpha \sqrt{\frac{m+n}{m \times n}}$, with $m$ and $n$ denoting the dimensions (number of events) of hypotheses $H_0$ and $H_1$, respectively. 
Here, $CDF_N(H)$ represents the cumulative distribution function of the normalized distribution corresponding to hypothesis $H$. The constant $c_\alpha$ for a given value of $\alpha$ is defined through
\begin{equation}\label{eq:calpha-kolmogorov}
1-\alpha = CDF_K(c_\alpha),
\end{equation}
where $CDF_K$ is the cumulative distribution function of the Kolmogorov distribution, given by
\begin{equation}\label{eq:cdf-kolmogorov}
CDF_K(x) = \dfrac{\sqrt{2\pi}}{x} \sum_{k=1}^{\infty} 
\exp\!\left[-\dfrac{(2k-1)^2\pi^2}{8x^2}\right].
\end{equation}
The values of $c_\alpha$ for a few representative choices of $\alpha$ are listed in Table~\ref{tab:alpha-calpha}.

\begin{table}[h!]
\caption{\label{tab:alpha-calpha} Values of $c_\alpha$ for selected values of $\alpha$ obtained using Eq.~(\ref{eq:calpha-kolmogorov}).}
	\renewcommand{\arraystretch}{1.50}
	\centering	
	\begin{tabular}{|c|c|c|c|c|c|c|} \hline
		$\alpha$ & $0.10$ & $0.05$ & $0.025$ & $0.01$ & $0.005$ & $0.001$ \\ \hline
		$c_\alpha$ & $1.22$ & $1.36$ & $1.48$ & $1.63$ & $1.73$ & $1.95$ \\ \hline
	\end{tabular}
\end{table}

The KS test is inherently a one-dimensional test and therefore requires a single variable. One could directly use $\Delta\phi_{j_1j_2}$, which exhibits the largest difference among the signal scenarios, while ignoring the remaining variables. 
However, in order to incorporate information from all relevant kinematic variables into the KS test, we construct a single optimal variable that maximally captures the discriminatory power of the full variable set. This is achieved using LDA.

\begin{figure}[ht!]
	\centering
	\includegraphics[width=0.45\textwidth]{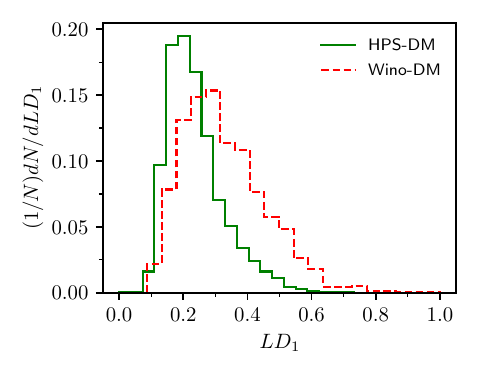}
	\includegraphics[width=0.45\textwidth]{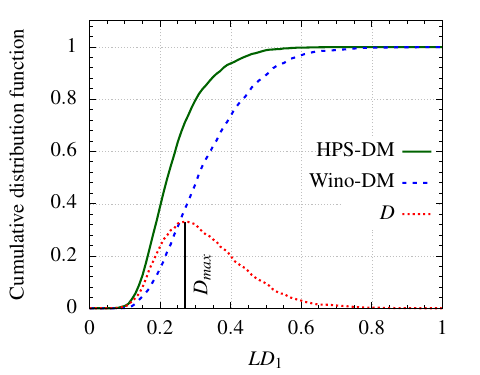}
	\includegraphics[width=0.45\textwidth]{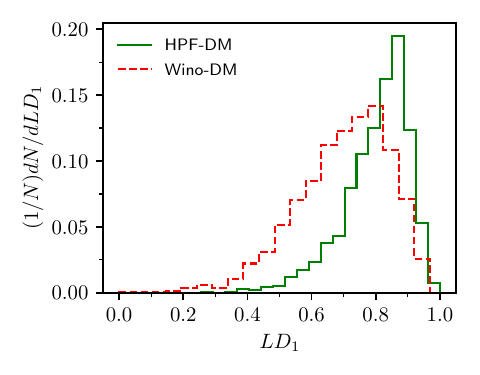}
	\includegraphics[width=0.45\textwidth]{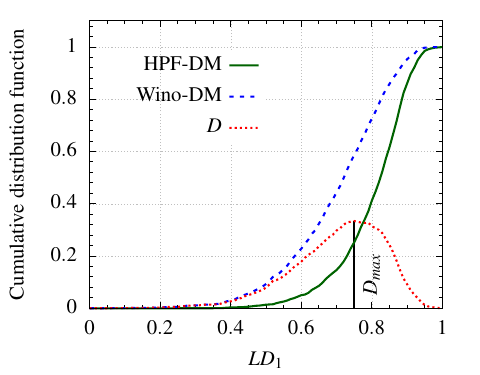}
	\includegraphics[width=0.45\textwidth]{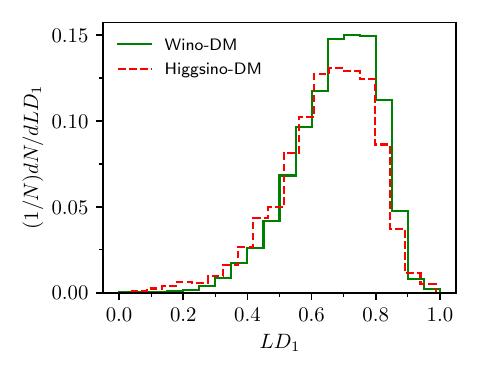}
	\includegraphics[width=0.45\textwidth]{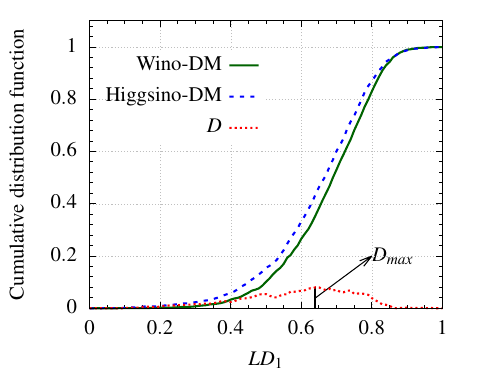}
	\caption{\label{fig:LDA-kolmogorov} Normalized distributions of the LDA variable $LD_1$, scaled to the range $[0,1]$, for pairs of signal scenarios (left column), and the corresponding cumulative distributions used in the KS test (right column).}
\end{figure}

The desired variable, denoted as $LD_1$, is obtained by projecting the input variable vector $\vec{X}$ onto the eigenvector corresponding to the largest eigenvalue of the matrix
\begin{equation}
S = S_W^{-1} S_B ,
\end{equation}
where
\begin{eqnarray}
S_W &=& \sum\limits_{i=1}^{\text{class}} (N_i - 1)\,\Sigma_i, \qquad
\Sigma_i = \frac{1}{N_i - 1} \sum\limits_{\vec{X} \in D_i} 
(\vec{X} - \vec{\mu}_i)(\vec{X} - \vec{\mu}_i)^T , \nonumber\\
S_B &=& \sum\limits_{i=1}^{\text{class}} N_i\,(\vec{\mu}_i - \vec{\mu})(\vec{\mu}_i - \vec{\mu})^T .
\end{eqnarray}
Here, $N_i$ is the number of events in the $i^{\text{th}}$ class, $D_i$ denotes the set of events in the $i^{\text{th}}$ class,  $\vec{\mu}_i$ is the mean vector of the $i^{\text{th}}$ class, and $\vec{\mu}$ is the overall mean vector obtained by combining all classes. 
The input variable vector $\vec{X}$ consists of the following  kinematic observables:
\begin{equation}\label{eq:obs-lda}
{p_T}_{j_1},~
{p_T}_{j_2},~
M_{j_1j_2},~
\cancel{E}_T,~
\Delta\eta_{j_1j_2},~
\Delta\phi_{j_1j_2},~
\Delta\phi_{j_1\cancel{E}_T},~
\Delta\phi_{j_2\cancel{E}_T}.
\end{equation}

The normalized distributions of the $LD_1$ variable, scaled to the range $[0,1]$, for pairs of signal scenarios are shown in the left column of Fig.~\ref{fig:LDA-kolmogorov}. The corresponding cumulative distributions used for the KS test are displayed in the right column. 
We perform this analysis for the signal pairs HPS--Wino, HPF--Wino, HPS--HPF, and Wino--Higgsino. 
The qualitative distinctions discussed earlier in connection with Figs.~\ref{fig:Dist-normal-1} and~\ref{fig:Dist-normal-2} are clearly reflected in Fig.~\ref{fig:LDA-kolmogorov}.

The resulting values of $\alpha$ (or equivalently, the $p$-values), obtained using Eq.~(\ref{eq:calpha-kolmogorov}), are presented in Table~\ref{tab:kolmogorov-result} for each signal pair. The table also lists the corresponding values of $D_{\rm max}$ and the maximum $c_\alpha$, with the number of events normalized to an integrated luminosity of ${\cal L}=3$ ab$^{-1}$. 
The HPS--Wino and HPF--Wino signal pairs can be distinguished with more than $99.99\%$ C.L. (corresponding to approximately $5\sigma$). 
The HPS and HPF signals are distinguishable at the $84\%$ C.L., while the Wino and Higgsino signals can be separated at the $90\%$ C.L.

\begin{table}
	\caption{\label{tab:kolmogorov-result} Results of the KS test for pairwise discrimination among the signal scenarios. The number of events is normalized to an integrated luminosity of ${\cal L}=3$ ab$^{-1}$.}
	\renewcommand{\arraystretch}{1.50}
	\centering	
	\begin{tabular}{|c|c|c|c|c|}\hline
		& HPS -- Wino & HPF -- Wino & HPS -- HPF & Wino -- Higgsino \\ \hline
		$D_{\rm max}$ & $0.331$ & $0.334$ & $0.0433$ & $0.0805$ \\ \hline
		$m,n$ & $1446,1420$ & $1300,1420$ & $1446,1300$ & $1420,280$ \\ \hline
		$c_\alpha|_{\rm max}$ & $8.860$ & $8.70$ & $1.128$ & $1.231$ \\ \hline
		$\alpha$ ($p$-value) & $0.0$ & $0.0$ & $0.157$ & $0.0965$ \\ \hline
		$(1-\alpha)$ C.L. & $100\%$ & $100\%$ & $84\%$ & $90\%$ \\ \hline
	\end{tabular}
\end{table}

To understand the relative contributions of the individual variables to the discrimination power reflected in Table~\ref{tab:kolmogorov-result}, we examine the components of the $LD_1$ variable. 
The $LD_1$ variable is expressed as a linear combination of the input observables listed in Eq.~(\ref{eq:obs-lda}), with the following mixing coefficients:
\begin{eqnarray}
\text{HPS -- Wino} &:& [0.609,\,1.63,\,0.0335,\,-0.0272,\,-98.9,\,9.13,\,8.18,\,8.33], \nonumber\\
\text{HPF -- Wino} &:& [-0.186,\,-0.887,\,0.0544,\,-0.149,\,99.7,\,-4.17,\,6.82,\,-1.08], \nonumber\\
\text{Wino -- Higgsino} &:& [-1.25,\,-0.289,\,-0.0102,\,0.902,\,37.9,\,-20.2,\,-81.9,\,37.7], \nonumber\\
\text{HPS -- HPF} &:& [-1.89,\,-0.381,\,0.107,\,1.93,\,-39.7,\,-11.5,\,-63.3,\,-65.4].
\end{eqnarray}
For the Higgs portal versus Wino discrimination, the variable $\Delta\eta_{j_1j_2}$ provides the dominant contribution ($\sim 99\%$). 
In the Wino versus Higgsino case, the variable $\Delta\phi_{j_1\cancel{E}_T}$ contributes most significantly ($\sim 82\%$). 
For the two Higgs portal scenarios, where the separability is comparatively weak, the variables $\Delta\phi_{j_1\cancel{E}_T}$ and $\Delta\phi_{j_2\cancel{E}_T}$ contribute almost equally.

\section{Conclusion}\label{sec:conclusion}
In this paper, we performed a comprehensive collider study of DM production via the VBF process at the future HL-LHC. Our primary objective was to distinguish between Higgs portal DM and neutralino DM scenarios, focusing on wino- and higgsino-like LSP signals in the VBF $2j + \ETmiss$ final state.

We observed that the polarization of weak gauge bosons in the VBF process plays a crucial role in shaping the kinematic distributions of the tagged jets. In particular, the transverse momentum of the jets is found to be softer in the Higgs portal scenario compared to the neutralino LSP scenarios. This behavior can be attributed to the dominance of longitudinally polarized $W$ bosons in the Higgs portal case, whereas transversely polarized $W$ bosons dominate the production of wino- and higgsino-like LSP pairs.

Additionally, significant differences were observed in the angular variables $\Delta\eta$ and $\Delta\phi$ of the two forward jets, which provide strong discriminating power between the Higgs portal and neutralino DM scenarios. In particular, the asymmetry $A_{\Delta\phi}$ exhibits opposite signs: it is negative for the Higgs portal scenario and positive for the neutralino LSP scenarios. While the two Higgs portal cases (scalar and fermionic DM) do not show substantial differences, mild but noticeable distinctions are observed between the wino- and higgsino-like neutralino signals.

To quantitatively assess the distinguishability among these scenarios, we performed a Kolmogorov--Smirnov  test. Our results indicate that the Higgs portal signals, both scalar (HPS) and fermionic (HPF), can be distinguished from the neutralino LSP signals with a C.L. exceeding $99.99\%$, corresponding to a $5\sigma$ significance. Furthermore, the HPS and HPF scenarios themselves can be differentiated at the $84\%$ C.L., while the wino- and higgsino-like neutralino signals show a separation at the $90\%$ C.L..

Among the kinematic observables considered, $\Delta\eta$ and $\Delta\phi_{j1\cancel{E}_T}$ play particularly important roles. The variable $\Delta\eta$ is highly effective in separating Higgs portal scenarios from neutralino DM, whereas $\Delta\phi_{j1\cancel{E}_T}$ provides significant discriminating power between wino- and higgsino-like neutralino signals.

In summary, our analysis demonstrates, as a proof of principle, that VBF processes at the HL-LHC offer a powerful and viable probe for discriminating between different dark matter candidates. The polarization structure of weak bosons, together with carefully chosen angular observables such as $\Delta\eta$ and $\Delta\phi$, provides a robust framework for collider-based discrimination between dark matter models, allowing us to distinguish Higgs portal DM from neutralino DM, as well as to probe the nature of the neutralino LSP itself.
\section*{Acknowledgments}
RR is supported by the FAPESP fellowships under grants 2023/04036-1 and 2025/06648-0. TG is grateful to acknowledge the support of the Harish-Chandra Research Institute and Homi Bhabha National Institute (HBNI), Mumbai. The authors are grateful to  Rohini  Godbole for insightful discussions during the early stages of this project. The authors dedicate this paper to her memory, honoring her profound contributions to high energy physics.
\bibliography{Ref}

\providecommand{\href}[2]{#2}\begingroup\raggedright\begin{thebibliography}{10}

\bibitem{Rubin:1970zza}
V.~C. Rubin and W.~K. Ford, Jr., {\it {Rotation of the Andromeda Nebula from a
  Spectroscopic Survey of Emission Regions}},
  \href{http://dx.doi.org/10.1086/150317}{{\em Astrophys. J.} {\bfseries 159}
  (1970) 379--403}.

\bibitem{Zwicky:1937zza}
F.~Zwicky, {\it {On the Masses of Nebulae and of Clusters of Nebulae}},
  \href{http://dx.doi.org/10.1086/143864}{{\em Astrophys. J.} {\bfseries 86}
  (1937) 217--246}.

\bibitem{Zwicky:1933gu}
F.~Zwicky, {\it {Die Rotverschiebung von extragalaktischen Nebeln}},
  \href{http://dx.doi.org/10.1007/s10714-008-0707-4}{{\em Helv. Phys. Acta}
  {\bfseries 6} (1933) 110--127}.

\bibitem{Hayashi:2006kw}
E.~Hayashi and S.~D.~M. White, {\it {How Rare is the Bullet Cluster?}},
  \href{http://dx.doi.org/10.1111/j.1745-3933.2006.00184.x}{{\em Mon. Not. Roy.
  Astron. Soc.} {\bfseries 370} (2006) L38--L41},
  \href{http://arxiv.org/abs/astro-ph/0604443}{{\ttfamily
  arXiv:astro-ph/0604443}}.

\bibitem{Clowe:2006eq}
D.~Clowe, M.~Bradac, A.~H. Gonzalez, M.~Markevitch, S.~W. Randall, C.~Jones,
  and D.~Zaritsky, {\it {A direct empirical proof of the existence of dark
  matter}},  \href{http://dx.doi.org/10.1086/508162}{{\em Astrophys. J. Lett.}
  {\bfseries 648} (2006) L109--L113},
  \href{http://arxiv.org/abs/astro-ph/0608407}{{\ttfamily
  arXiv:astro-ph/0608407}}.

\bibitem{Hu:2001bc}
W.~Hu and S.~Dodelson, {\it {Cosmic Microwave Background Anisotropies}},
  \href{http://dx.doi.org/10.1146/annurev.astro.40.060401.093926}{{\em Ann.
  Rev. Astron. Astrophys.} {\bfseries 40} (2002) 171--216},
  \href{http://arxiv.org/abs/astro-ph/0110414}{{\ttfamily
  arXiv:astro-ph/0110414}}.

\bibitem{WMAP:2006bqn}
{\bfseries WMAP} Collaboration, D.~N. Spergel {\em et~al.}, {\it {Wilkinson
  Microwave Anisotropy Probe (WMAP) three year results: implications for
  cosmology}},  \href{http://dx.doi.org/10.1086/513700}{{\em Astrophys. J.
  Suppl.} {\bfseries 170} (2007) 377},
  \href{http://arxiv.org/abs/astro-ph/0603449}{{\ttfamily
  arXiv:astro-ph/0603449}}.

\bibitem{WMAP:2012nax}
{\bfseries WMAP} Collaboration, G.~Hinshaw {\em et~al.}, {\it {Nine-Year
  Wilkinson Microwave Anisotropy Probe (WMAP) Observations: Cosmological
  Parameter Results}},
  \href{http://dx.doi.org/10.1088/0067-0049/208/2/19}{{\em Astrophys. J.
  Suppl.} {\bfseries 208} (2013) 19},
  \href{http://arxiv.org/abs/1212.5226}{{\ttfamily arXiv:1212.5226
  [astro-ph.CO]}}.

\bibitem{Planck:2018vyg}
{\bfseries Planck} Collaboration, N.~Aghanim {\em et~al.}, {\it {Planck 2018
  results. VI. Cosmological parameters}},
  \href{http://dx.doi.org/10.1051/0004-6361/201833910}{{\em Astron. Astrophys.}
  {\bfseries 641} (2020) A6}, \href{http://arxiv.org/abs/1807.06209}{{\ttfamily
  arXiv:1807.06209 [astro-ph.CO]}}. [Erratum: Astron.Astrophys. 652, C4
  (2021)].

\bibitem{Baek:2011aa}
S.~Baek, P.~Ko, and W.-I. Park, {\it {Search for the Higgs portal to a singlet
  fermionic dark matter at the LHC}},
  \href{http://dx.doi.org/10.1007/JHEP02(2012)047}{{\em JHEP} {\bfseries 02}
  (2012) 047}, \href{http://arxiv.org/abs/1112.1847}{{\ttfamily arXiv:1112.1847
  [hep-ph]}}.

\bibitem{CMS:2018tip}
{\bfseries CMS} Collaboration, {\it {Search for invisible decays of a Higgs
  boson produced through vector boson fusion at the High-Luminosity LHC}},
  {\em CMS-PAS-FTR-18-016} (2018) .
  \url{https://cds.cern.ch/record/2647700/files/FTR-18-016-pas.pdf}.

\bibitem{Cepeda:2019klc}
M.~Cepeda {\em et~al.}, {\it {Report from Working Group 2}: {Higgs Physics at
  the HL-LHC and HE-LHC}},
  \href{http://dx.doi.org/10.23731/CYRM-2019-007.221}{{\em CERN Yellow Rep.
  Monogr.} {\bfseries 7} (2019) 221--584},
  \href{http://arxiv.org/abs/1902.00134}{{\ttfamily arXiv:1902.00134
  [hep-ph]}}.

\bibitem{Knuth:10.5555/270146}
D.~E. Knuth, {\em The Art of Computer Programming, Volume 2 (3rd Ed.):
  Seminumerical Algorithms}.
\newblock Addison-Wesley Longman Publishing Co., Inc., USA, 1997.

\bibitem{pratt2012concepts}
J.~Pratt and J.~Gibbons, {\em Concepts of Nonparametric Theory}.
\newblock Springer Series in Statistics. Springer New York, 2012.
\newblock \url{https://books.google.co.in/books?id=PCHnBwAAQBAJ}.

\bibitem{tharwat:2017}
A.~Tharwat, T.~Gaber, A.~Ibrahim, and A.~E. Hassanien, {\it Linear discriminant
  analysis: A detailed tutorial},
  \href{http://dx.doi.org/10.3233/AIC-170729}{{\em Ai Communications}
  {\bfseries 30} (05, 2017) 169--190,}.

\bibitem{ghojogh2019linear}
B.~Ghojogh and M.~Crowley, {\it Linear and quadratic discriminant analysis:
  Tutorial},  \href{http://arxiv.org/abs/1906.02590}{{\ttfamily
  arXiv:1906.02590 [stat.ML]}}.

\bibitem{Profumo:2007wc}
S.~Profumo, M.~J. Ramsey-Musolf, and G.~Shaughnessy, {\it {Singlet Higgs
  phenomenology and the electroweak phase transition}},
  \href{http://dx.doi.org/10.1088/1126-6708/2007/08/010}{{\em JHEP} {\bfseries
  08} (2007) 010}, \href{http://arxiv.org/abs/0705.2425}{{\ttfamily
  arXiv:0705.2425 [hep-ph]}}.

\bibitem{ATLAS:2015ciy}
{\bfseries ATLAS} Collaboration, G.~Aad {\em et~al.}, {\it {Constraints on new
  phenomena via Higgs boson couplings and invisible decays with the ATLAS
  detector}},  \href{http://dx.doi.org/10.1007/JHEP11(2015)206}{{\em JHEP}
  {\bfseries 11} (2015) 206}, \href{http://arxiv.org/abs/1509.00672}{{\ttfamily
  arXiv:1509.00672 [hep-ex]}}.

\bibitem{ATLAS:2016neq}
{\bfseries ATLAS, CMS} Collaboration, G.~Aad {\em et~al.}, {\it {Measurements
  of the Higgs boson production and decay rates and constraints on its
  couplings from a combined ATLAS and CMS analysis of the LHC pp collision data
  at $ \sqrt{s}=7 $ and 8 TeV}},
  \href{http://dx.doi.org/10.1007/JHEP08(2016)045}{{\em JHEP} {\bfseries 08}
  (2016) 045}, \href{http://arxiv.org/abs/1606.02266}{{\ttfamily
  arXiv:1606.02266 [hep-ex]}}.

\bibitem{Robens:2015gla}
T.~Robens and T.~Stefaniak, {\it {Status of the Higgs Singlet Extension of the
  Standard Model after LHC Run 1}},
  \href{http://dx.doi.org/10.1140/epjc/s10052-015-3323-y}{{\em Eur. Phys. J. C}
  {\bfseries 75} (2015) 104}, \href{http://arxiv.org/abs/1501.02234}{{\ttfamily
  arXiv:1501.02234 [hep-ph]}}.

\bibitem{Kanemura:2010sh}
S.~Kanemura, S.~Matsumoto, T.~Nabeshima, and N.~Okada, {\it {Can WIMP Dark
  Matter overcome the Nightmare Scenario?}},
  \href{http://dx.doi.org/10.1103/PhysRevD.82.055026}{{\em Phys. Rev. D}
  {\bfseries 82} (2010) 055026},
  \href{http://arxiv.org/abs/1005.5651}{{\ttfamily arXiv:1005.5651 [hep-ph]}}.

\bibitem{ATLAS:2019lng}
{\bfseries ATLAS} Collaboration, G.~Aad {\em et~al.}, {\it {Searches for
  electroweak production of supersymmetric particles with compressed mass
  spectra in $\sqrt{s}=$ 13 TeV $pp$ collisions with the ATLAS detector}},
  \href{http://dx.doi.org/10.1103/PhysRevD.101.052005}{{\em Phys. Rev. D}
  {\bfseries 101} no.~5, (2020) 052005},
  \href{http://arxiv.org/abs/1911.12606}{{\ttfamily arXiv:1911.12606
  [hep-ex]}}.

\bibitem{ATLAS:2021moa}
{\bfseries ATLAS} Collaboration, G.~Aad {\em et~al.}, {\it {Search for
  chargino\textendash{}neutralino pair production in final states with three
  leptons and missing transverse momentum in $\sqrt{s} = 13$~TeV pp collisions
  with the ATLAS detector}},
  \href{http://dx.doi.org/10.1140/epjc/s10052-021-09749-7}{{\em Eur. Phys. J.
  C} {\bfseries 81} no.~12, (2021) 1118},
  \href{http://arxiv.org/abs/2106.01676}{{\ttfamily arXiv:2106.01676
  [hep-ex]}}.

\bibitem{ATLAS:2024umc}
{\bfseries ATLAS} Collaboration, G.~Aad {\em et~al.}, {\it {Search for Nearly
  Mass-Degenerate Higgsinos Using Low-Momentum Mildly Displaced Tracks in pp
  Collisions at s=13\,\,TeV with the ATLAS Detector}},
  \href{http://dx.doi.org/10.1103/PhysRevLett.132.221801}{{\em Phys. Rev.
  Lett.} {\bfseries 132} no.~22, (2024) 221801},
  \href{http://arxiv.org/abs/2401.14046}{{\ttfamily arXiv:2401.14046
  [hep-ex]}}.

\bibitem{ATLAS:2025lhc}
{\bfseries ATLAS} Collaboration, G.~Aad {\em et~al.}, {\it {Search for
  higgsinos in compressed mass spectra using low-momentum tracks in $pp$
  collisions at $\sqrt{s}=13$ TeV with the ATLAS detector}},
  \href{http://arxiv.org/abs/2511.20042}{{\ttfamily arXiv:2511.20042
  [hep-ex]}}.

\bibitem{Dawson:1984gx}
S.~Dawson, {\it {The Effective W Approximation}},
  \href{http://dx.doi.org/10.1016/0550-3213(85)90038-0}{{\em Nucl. Phys. B}
  {\bfseries 249} (1985) 42--60}.

\bibitem{Kane:1984bb}
G.~L. Kane, W.~W. Repko, and W.~B. Rolnick, {\it {The Effective W+-, Z0
  Approximation for High-Energy Collisions}},
  \href{http://dx.doi.org/10.1016/0370-2693(84)90105-9}{{\em Phys. Lett. B}
  {\bfseries 148} (1984) 367--372}.

\bibitem{Borel:2012by}
P.~Borel, R.~Franceschini, R.~Rattazzi, and A.~Wulzer, {\it {Probing the
  Scattering of Equivalent Electroweak Bosons}},
  \href{http://dx.doi.org/10.1007/JHEP06(2012)122}{{\em JHEP} {\bfseries 06}
  (2012) 122}, \href{http://arxiv.org/abs/1202.1904}{{\ttfamily arXiv:1202.1904
  [hep-ph]}}.

\bibitem{Cornwall:1974km}
J.~M. Cornwall, D.~N. Levin, and G.~Tiktopoulos, {\it {Derivation of Gauge
  Invariance from High-Energy Unitarity Bounds on the s Matrix}},
  \href{http://dx.doi.org/10.1103/PhysRevD.10.1145}{{\em Phys. Rev. D}
  {\bfseries 10} (1974) 1145}. [Erratum: Phys.Rev.D 11, 972 (1975)].

\bibitem{Kuroda:1999ks}
M.~Kuroda, {\it {Complete Lagrangian of MSSM}},
  \href{http://arxiv.org/abs/hep-ph/9902340}{{\ttfamily arXiv:hep-ph/9902340}}.

\bibitem{CMS:2016dhk}
{\bfseries CMS} Collaboration, V.~Khachatryan {\em et~al.}, {\it {Searches for
  invisible decays of the Higgs boson in pp collisions at $\sqrt{s}$ = 7, 8,
  and 13 TeV}},  \href{http://dx.doi.org/10.1007/JHEP02(2017)135}{{\em JHEP}
  {\bfseries 02} (2017) 135}, \href{http://arxiv.org/abs/1610.09218}{{\ttfamily
  arXiv:1610.09218 [hep-ex]}}.

\bibitem{Alwall:2014hca}
J.~Alwall, R.~Frederix, S.~Frixione, V.~Hirschi, F.~Maltoni, O.~Mattelaer,
  H.-S. Shao, T.~Stelzer, P.~Torrielli, and M.~Zaro, {\it The automated
  computation of tree-level and next-to-leading order differential cross
  sections, and their matching to parton shower simulations},
  \href{http://dx.doi.org/10.1007/JHEP07(2014)079}{{\em JHEP} {\bfseries 07}
  (2014) 079}, \href{http://arxiv.org/abs/1405.0301}{{\ttfamily arXiv:1405.0301
  [hep-ph]}}.

\bibitem{Sjostrand:2014zea}
T.~Sj{\"o}strand, S.~Ask, J.~R. Christiansen, R.~Corke, N.~Desai, P.~Ilten,
  S.~Mrenna, S.~Prestel, C.~O. Rasmussen, and P.~Z. Skands, {\it An
  introduction to {PYTHIA 8.2}},
  \href{http://dx.doi.org/10.1016/j.cpc.2015.01.024}{{\em Comput. Phys.
  Commun.} {\bfseries 191} (2015) 159},
\href{http://arxiv.org/abs/1410.3012}{{\ttfamily arXiv:1410.3012 [hep-ph]}}.

\bibitem{deFavereau:2013fsa}
{\bfseries DELPHES 3} Collaboration, J.~de~Favereau, C.~Delaere, P.~Demin,
  A.~Giammanco, V.~Lema{\^\i}tre, A.~Mertens, and M.~Selvaggi, {\it {DELPHES 3,
  A modular framework for fast simulation of a generic collider experiment}},
  \href{http://dx.doi.org/10.1007/JHEP02(2014)057}{{\em JHEP} {\bfseries 02}
  (2014) 057}, \href{http://arxiv.org/abs/1307.6346}{{\ttfamily arXiv:1307.6346
  [hep-ex]}}.

\end{thebibliography}\endgroup
\bibliographystyle{utphysM}

\end{document}